\documentclass{bmcart}

\usepackage{amsthm,amsmath}
\RequirePackage{natbib}
\usepackage[utf8]{inputenc} 
\usepackage{graphicx}
\usepackage{caption}

\startlocaldefs
\endlocaldefs

\begin{document}

\begin{frontmatter}

\begin{fmbox}
\dochead{Research}


\title{Brain Signals Analysis Based Deep Learning Methods: Recent advances in the study of non-invasive brain signals}


%
%

\author[
   addressref={aff1},                   
   email={a.essa@csuohio.edu}   
]{\inits{AE}\fnm{Almabrok} \snm{Essa}}
\author[
]{\inits{VA}\fnm{Hari} \snm{Kotte}}


\address[id=aff1]{
  \orgname{Department of Electrical Engineering and Computer Science, Cleveland State University}, 
  \street{2121 Euclid Ave},                     %
  \postcode{44115}                                
  \city{Cleveland, OH},                              
  \cny{USA}                                    
}  




\end{fmbox}


\begin{abstractbox}

\begin{abstract} 
Brain signals constitute the information that are processed by millions of brain neurons (nerve cells and brain cells). These brain signals can be recorded and analyzed using various of non-invasive techniques such as the Electroencephalograph (EEG), Magneto-encephalograph (MEG) as well as brain-imaging techniques such as Magnetic Resonance Imaging (MRI), Computed Tomography (CT) and others, which will be discussed briefly in this paper. This paper discusses about the currently emerging techniques such as the usage of different Deep Learning (DL) algorithms for the analysis of these brain signals and how these algorithms will be helpful in determining the neurological status of a person by applying the signal decoding strategy. 

%
\end{abstract}




\end{abstractbox}
%

\end{frontmatter}


\section{Introduction}
Brain signals measure the instinct biometric information from the human brain, which reflects the user’s passive or active mental state. Brain signals model the information that are processed by millions of neurons present in the brain in the form of signals. These brain signals resemble the neural activity (that includes both the sensory and motor activities) of the person. The sensory and motor activities of the person (or user) can be known upon the processing the brain signals. With the emerging technologies, the brain signals are analysed and processed using different traditional (EEG, MEG, MRI, fMRI) and non-traditional signal processing techniques (Deep Learning algorithms, Decision Tress, etc.). 

All traditional versions of the brain signal analysis include feature extraction step followed by the classification process at some point. Jahankhani et al. have conducted experiments and used Discrete Wavelet Transform (DWT) as a feature extraction technique for extracting the features from the EEG brain signals and multilayer perceptron is the classification technique that has been used along with the radial basis function network (RBF) \cite{jahankhani2006eeg} which has resulted in accurate EEG signal classification, in terms of training performance. While Acharya et al. have conducted experiments using the EEG signals along with the wavelet packet transform (WPT) as the feature extraction method and support vector machine (SVM) as the classification method \cite{acharya2011automatic}. These methods together resulted in the accurate detection of the epilepsy (a neurological disorder). Similarly, more other feature extraction techniques could be used that combine two or more different types of techniques like \cite{essa8, essa9}, which could be a potential fed to machine learning techniques. 

On the other hand, Williamson et al. have conducted experiments using machine learning algorithms that involves usage of multivariate EEG features. The delayed EEG data is computed with the help of the machine learning algorithms along with the help of support vector machine (SVM) that result in a prediction score for the seizure patients \cite{williamson2012seizure}. The seizure patients’ risk is predicted with the help of the correlation structure which was developed using numerous variations that have been modeled using the available EEG signal data. There are other feature extraction techniques have been applied such as phase and a lag synchronization measure to a selected subset of multi-contact intracranial EEG \cite{winterhalder2006spatio}. Similarly, more other feature extraction techniques could be used that may apply encoding and decoding (IED) strategy \cite{essa1, essa2} before feeding into the machine learning techniques.  

This paper discusses about what a brain computer interface (BCI) system is and what are the different functional components associated with the BCI system. Alos, it explains about each functional component of the BCI system by discussing the different ways of acquiring the brain signals and how these signals be processed. In addition, it discusses about the different classification methods that are used in the past and the current trending methods like using the deep learning algorithms for feature extraction and classification process. 

\section{Brain-Computer Interface (BCI)}
\begin{figure}[h!]
	\centering
	\includegraphics[width=0.85\textwidth]{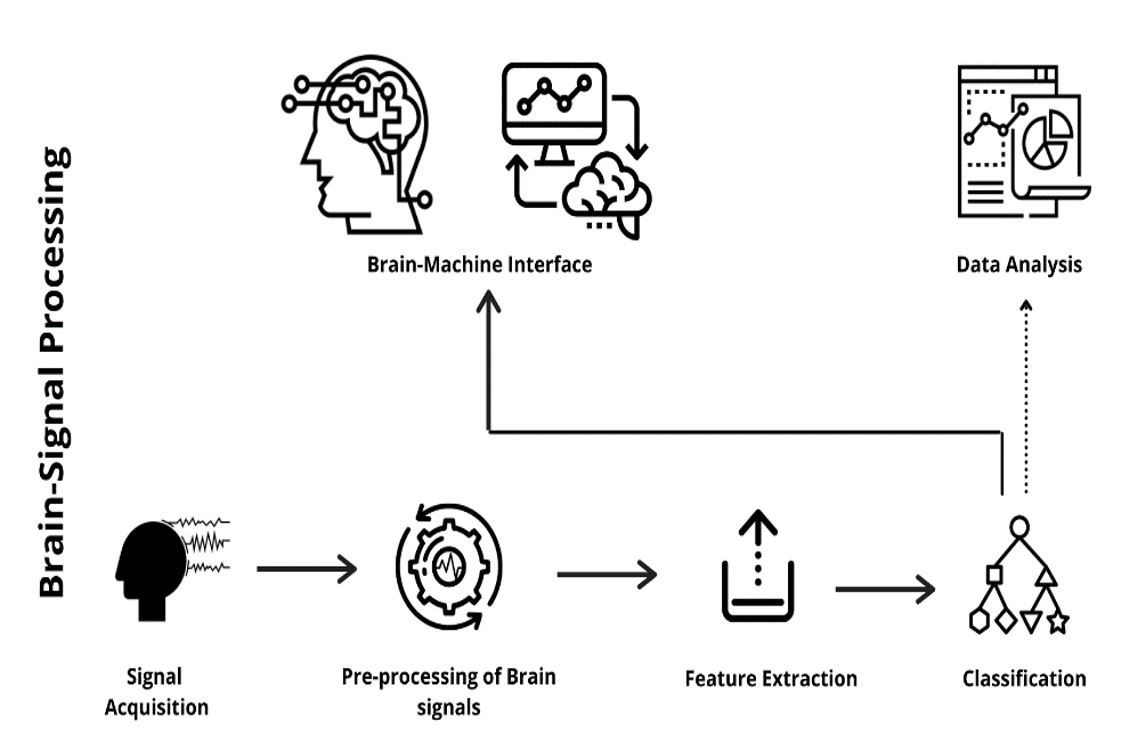}
	\caption{End-to-end brain signal analysis system, which known as a Brain-Computer Interface (BCI) system when the classified signals are applied to control smart equipment.}
	\label{fig1}
\end{figure}

Brain-computer Interface (BCI) uses the acquired brain signals to control the outside world smart devices. In orther to do so, these brain signals are further processed using different methods such as the Fast Fourier Transform (FFT), Wavelet Transform (WT), Time-Frequency Distribution (TFD), and more others. Upon processing these signals, the signals are sent further for classification using Artificial Neural Networks (ANNs), Support Vector Machine (SVM), Logistic Regression (LR) and other higher levels such as  Deep-Learning and Machine Learning (ML) algorithms \cite{shih2012brain}. The typical Brain-computer Interface (BCI) system along with its functional components is represented in Fig. \ref{fig1}.

\subsection{Brain Signal Acquisition}
Brain signals exist in five different waveforms, when recorded. The well-known five types of brain waveforms are alpha, beta, gamma, theta and delta waves based on the activity being performed by the subject and based on their amplitude and frequencies, which can be seen in Table \ref{table1} \cite{erman2009disorders}. These signals are acquired using electrodes utilizing invasive or non-invasive procedures.

\begin{table}[ht]
	\centering
	\caption{Different Brain waveforms of different frequencies during different activities \cite{erman2009disorders}.} 
	\label{table1}
	\includegraphics[width=12.75cm,height=6.7cm]{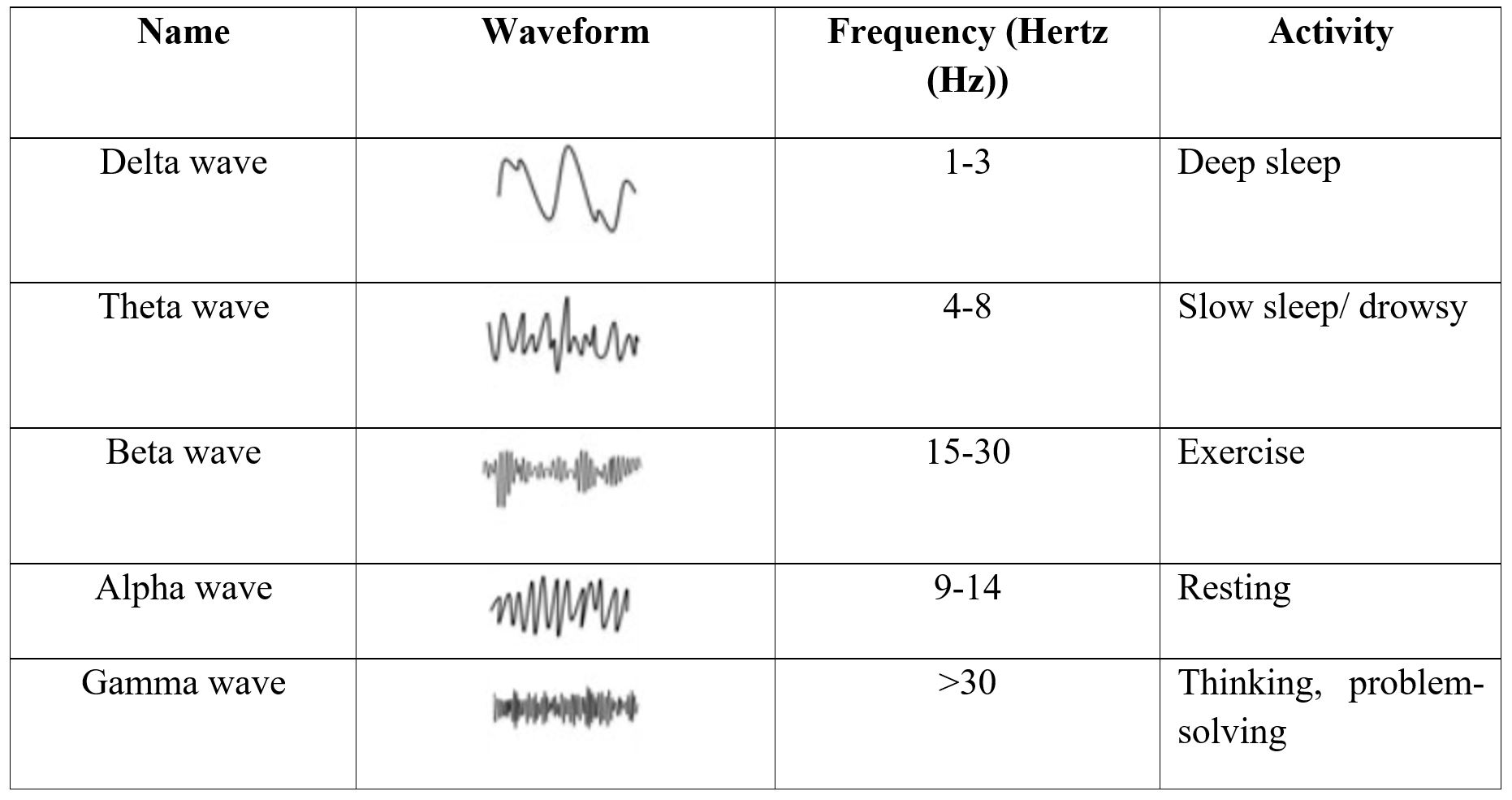}
\end{table}

There are different electrodes that are used for acquiring the signals. They are the needle electrodes and the surface or plate electrodes that are placed invasively and non-invasively on the surface of the brain respectively \cite{delgado1952permanent}. The needle electrodes are used when collecting the signal information from the interior regions of the brain while the surface electrodes are used when collecting the information from the surface of the brain. The needle electrodes do not follow any positioning system while the surface electrodes follow the 10-20 placement system that appears in the form a cap.  

\subsection*{Invasive Procedures}
\begin{figure}[h!]
	\centering
	\includegraphics[width=0.9\textwidth]{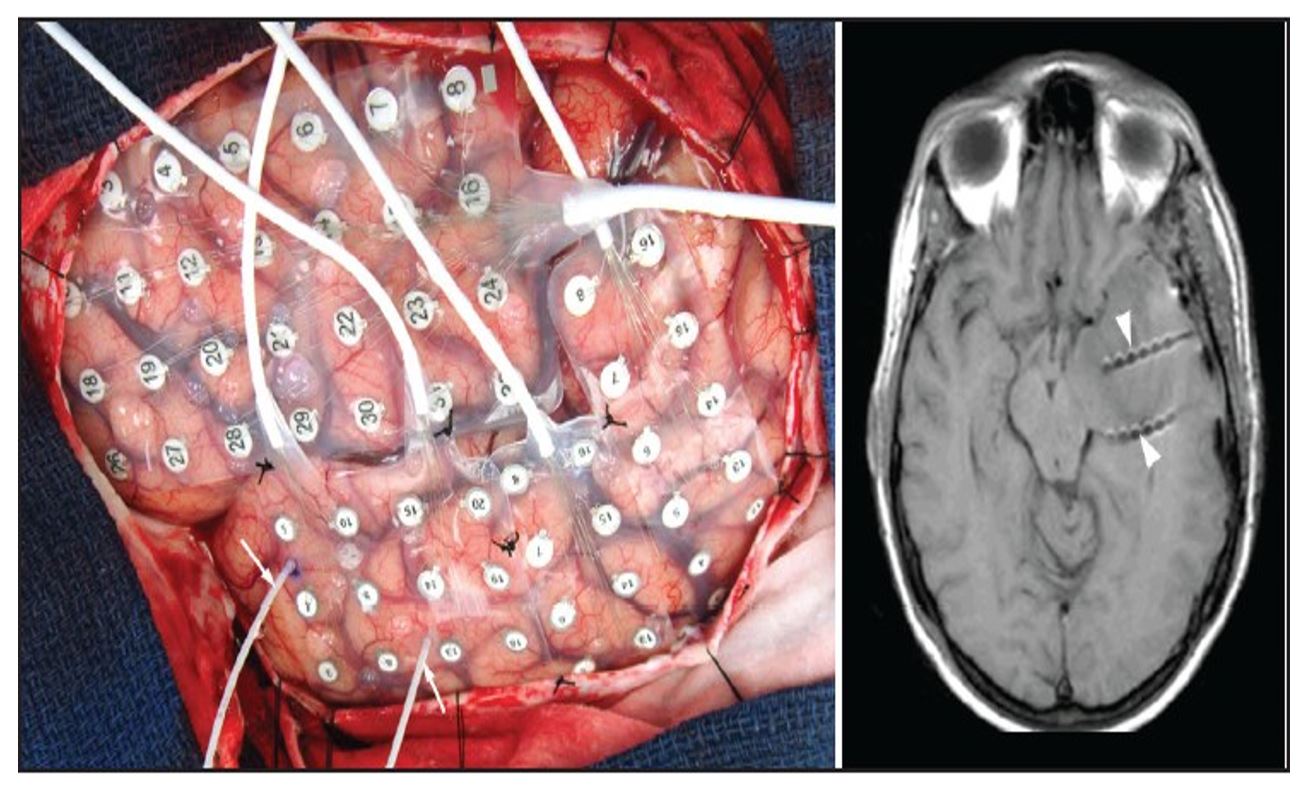}
	\caption{Electrocorticography (ECoG). Left, is an intraoperative photograph showing left cerebral hemisphere covered by various subdural grid electrode arrays. Right, a T1W axial magnetic resonance imaging. \cite{shah2014invasive}.}
	\label{fig2}
\end{figure}

There are few invasive imaging procedures that are used for monitoring the electrical activity of the brain. The most commonly used invasive procedures in the clinics and for the laboratory purpose are the Electrocorticography (ECoG), Deep neural stimulation, Invasive electro-encephalography (EEG), implantable electrodes and more. Electrocorticography (ECoG), also referred to as intracranial electroencephalography (iEEG) uses an array of electrodes as shown in Fig. \ref{fig2} that is placed directly on the surface of the soft tissue of the brain \cite{shah2014invasive, Nuwer}. This type of invasive procedure is used in identifying the epileptic regions and is also used while performing surgery, to evaluate the activity of the brain at the time of surgery.

\begin{figure}[h!]
	\centering
	\includegraphics[width=0.93\textwidth]{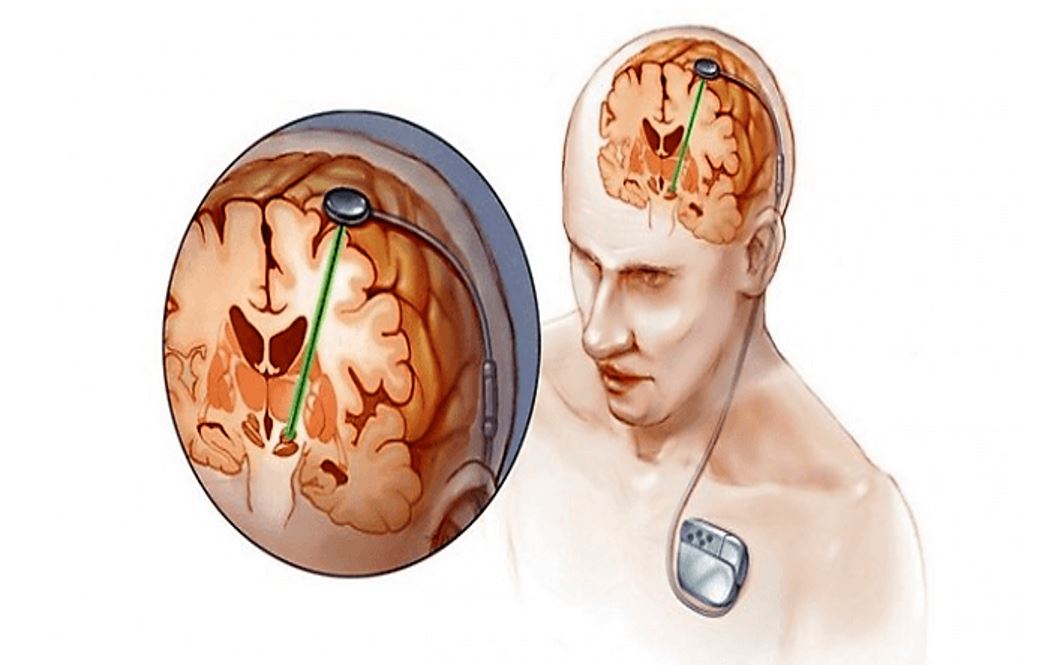}
	\caption{Deep Neural Stimulation (Electrode – surface of the brain, Pacemaker-under the clavicle) \cite{Nuwer}.}
	\label{fig3}
\end{figure}

Deep Neural Stimulation uses 2 electrodes (generally referred to as the anode and cathode) which are placed on the surface of the brain at specific areas. These electrodes are placed at a location which is responsible for the movement of the body parts (hands and legs). In deep neural stimulation, the electrodes that are connected to a pacemaker (conducts current to the ends of the electrodes through the wires) as can be seen in Fig. \ref{fig3}, that conduct certain amplitude of electric current to restore the activity in the specific areas by stimulating the neurons in that area. Deep Neural Stimulation is used in case of providing the temporary treatment to the person with neural disorders such as the Parkinson’s disease. Similarly to the heart signal acquisition \cite{essa20, essa21}, the brain signals need tools and procedures to be acquired. However, it would be a great deal of attention if the brain signals could be collected wireless as the heart signals \cite{essa22}.   

\subsection*{Non-Invasive Procedures}
The non-invasive methods of acquiring the data of the functioning of the brain can be done using different imaging modalities as mentioned in this section.
Brain imaging techniques are used in both diagnosing and treating the brain disorders. There are different Brain-imaging techniques that fall widely into two categories, which are either the structural imaging, in which the anatomy of brain can be studied or the functional imaging, in which the physiology or in other terms the functioning of the brain can be studied.
Electroencephalogram (EEG), Magnetoencephalogram (MEG), Magnetic Resonance Imaging (MRI), Computed Tomography (CT), Positron Emission tomography (PET), and function magnetic resonance imaging (fMRI) are few brain imaging techniques that are commonly used. Where the Computed Tomography (CT), Magnetic Resonance Imaging (MRI), functional Magnetic Resonance Imaging (fMRI) and Positron Emission Tomography (PET) are few imaging modalities that are used for detecting any abnormalities that are associated with the brain and its blood vessels. These imaging modalities are useful in diagnosing the changes in the brain. Where as the Computed Tomography (CT) is an imaging modality that uses X-rays for diagnostic/ clinical purpose. It is useful in detecting and helps the physician in identification of the abnormality associated with the brain.
Positron Emission Tomography (PET) is an imaging modality that is used in clinics for identifying disorders related to the organs and tissues  \cite{khoo1997magnetic}. It is different from the other imaging modalities because of its usage of traces of the radioactive elements (or tracers) in the form of a dye according to the national institue of health (NIH, 2021). These tracers help in diagnosing the location of the disease by coating colour to the diseased area or region.
Magnetic Resonance Imaging (MRI) is another imaging modality which is used to identify the anatomical changes of the brain by producing images of the brain in numerous slices according to the "Radiological Society of North America (RSNA) and American College of Radiology (ACR), “RMN functional (RMNf) (Radiologyinfo, 2021)" and lets the neurologist or the related medical staff in diagnosing the presence of a tumour or lesion whereas the functional MRI (fMRI) is used in producing the images of blood vessels of the brain and the blood circulating within these blood vessels \cite{khandpur1987handbook}. 



\subsubsection{10-20 Placement System}
\begin{figure}[h!]
	\centering
	\includegraphics[width=0.95\textwidth]{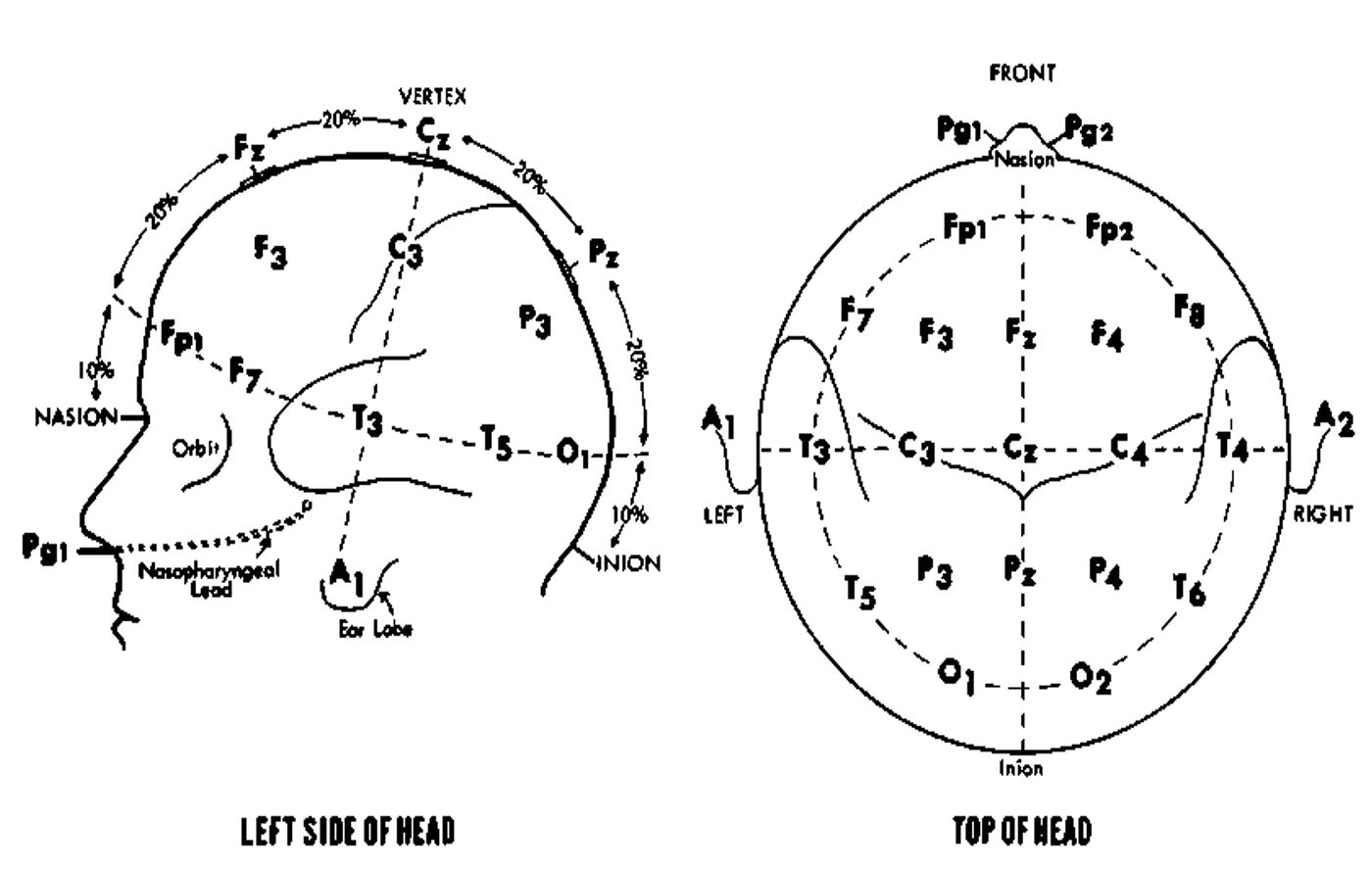}
	\caption{10-20 Electrode Placement system \cite{oxley2016minimally}.}
	\label{fig5}
\end{figure}

In an 10-20 system the electrodes are placed at 21 different locations on the scalp at a distance of 5\%, 10\% and 20\%. The electrodes that are placed on the side of the cranium are at a distance of 5\% whereas the other electrodes that are placed in different planes (coronal, sagittal planes) are placed at a distance of 10\% and 20\%. These 21 electrodes are named based on their position on the scalp as F (frontal), P (parietal), T (temporal), O (Occipital), C (central) and Fp (Frontal -polar) \cite{oxley2016minimally}. The electrodes are placed on the subjects’ scalp as explained in Fig. \ref{fig5}.

These signals that are acquired through electrodes are further processed using different or any of the existing non-invasive procedures such as the Electroencephalography (EEG), Magnetoencephalography (MEG) and Evoked potentials (EPs) \cite{khandpur1987handbook}. There are other non-invasive techniques such as the Electrooculography (EOG), Steady-state visual Evoked potentials (SSVEPs), etcetera that use visual cortex and pupil movements as reference for acquiring and processing the signals.

\subsubsection{Electroencephalography (EEG)}
\begin{figure}[h!]
	\centering
	\includegraphics[width=1\textwidth]{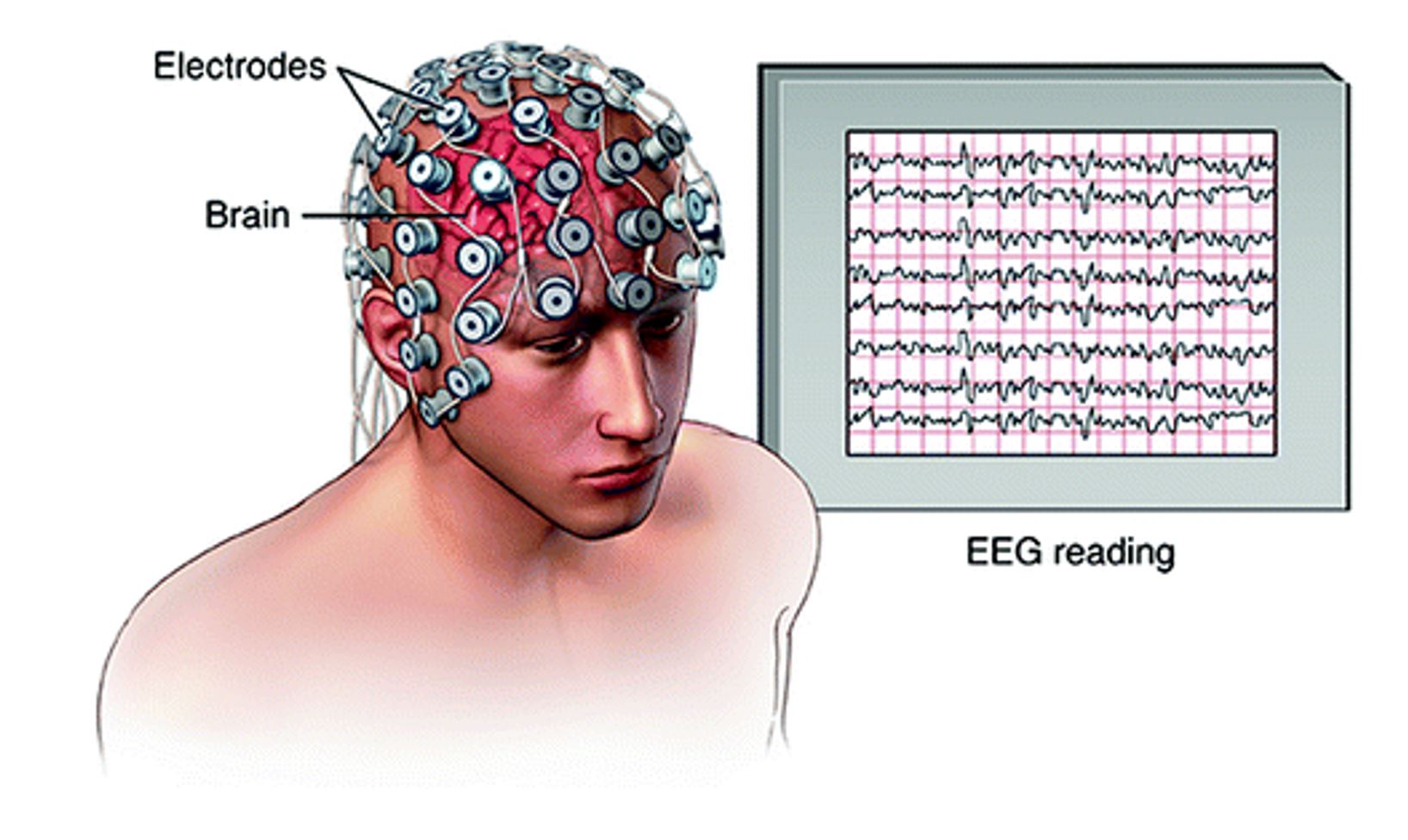}
	\caption{Electroencephalography \cite{siuly2016electroencephalogram}. The monitor shows the Encephalogram (EEG) which shows the non-overlapping brain signals that are associated with different brain activities.}
	\label{fig6}
\end{figure}

Electroencephalography (EEG) is a technique of recording the brain signals with the help of the surface electrodes placed on the scalp and using the 10-20 placement system. In the past decades, the brain signals are recorded traditionally (as analogue) and processed traditionally. With the advancing technology, the traditional equipment is replaced by the digital medical systems with a built-in filters and amplifiers for the signals \cite{binnie1994electroencephalography}.
Electroencephalography is mainly used for studying the brain activities and used for diagnostic purposes (for identifying the abnormal activities of the brain). Fig. \ref{fig6} shows the components involved in an electroencephalographic procedure (the electrodes and the monitor that shows the recorded brain signals).

\subsubsection{Magnetoencephalography (MEG)}
Magnetoencephalography (MEG) is a developing neuroimaging technique that maps all around the brain based on the magnetic fields, which are produced as a result of the electric activities of the brain. The MEG signals that are recorded by the brain at different magnetic fields refer to the different activities associated with the brain \cite{hamalainen1993magnetoencephalography}. These magnetic fields are identified by magnetometers present in a Magnetoencephalography machine. Magnetoencephalography is more preferred these days over the EEG because of its spatial resolution.

\subsubsection{Evoked Potential (EP)}
Evoked potentials are the result of the sensory triggers or in other words they are the result of the electrical activity in the sensory areas of the brain. Evoked potentials are seen in an electroencephalogram (EEG) and in order to acquire just the recorded activity of the evoked potentials from the EEG, the signals should be filtered out \cite{sarma2016pre}.

\subsection{Pre-Processing and Feature Extraction}
The pre-processing and feature extraction play an important role in the analysis and processing of the acquired brain signals. They are necessary components for efficient and accurate brain based diagnosis.   

\subsubsection{Pre-processing}
The brain signals that are acquired from the subject or through the electrodes are further processed by a data acquisition system or a data acquisition software, where these signals of small amplitudes and frequencies (about $1-20 Hz$) are amplified and filtered. In the pre-processing of the brain signals, the unnecessary artifacts, and effects of noise on the signal is removed and reduced, respectively. There are different techniques used for the pre-processing like using filters of different frequencies. For example, using a notch filter will help the user in reducing the effects of the noise on the acquired signals \cite{wang2011eeg}. While using a low-pass or high-pass filter or while using a band-pass filter or a band-reject filter, certain level of frequency is set to avoid the extra- width (or frequency) of the signal.

\subsubsection{Feature Extraction}
Feature extraction involves different techniques considering the signals with respect to the time, frequency and/or time-frequency domains which can be seen in Table \ref{table2}. Feature extraction is used to acquire the features that are present in the acquired brain signals. Traditional machine learning approaches utilize handmade features by applying several feature extraction algorithms including Discrete Fourier Transform (DFT), Fast Fourier Transform (FFT), Autoregression, Mean, Median, Standard Deviation (SD), Discrete Cosine Transform (DCT), as well as other techniques that have been have been used and proven highe performance in other application such as local edge/corner feature integration (LFI) \cite{essa4}, local boosted features (LBF) \cite{essa6}, Eigenvectors , and Discrete Wavelet Transform (DWT) \cite {al2014methods}. Also, there are some methods that combine two or more different types of techniques as in \cite{essa5, essa3, ubeyli2008analysis, subasi2005neural, essa7}. However, in the case of deep learning, the features are represented hierarchically in multiple levels and learned automatically, which is the key difference in feature extraction between traditional machine learning and deep learning techniques. 

\begin{table}[h]
	\centering
	\caption{Feature Extraction parameters as a function of time and frequency \cite{liang2012automatic, amin2017classification}.} 
	\label{table2}
	\includegraphics[width=12.5cm,height=12cm]{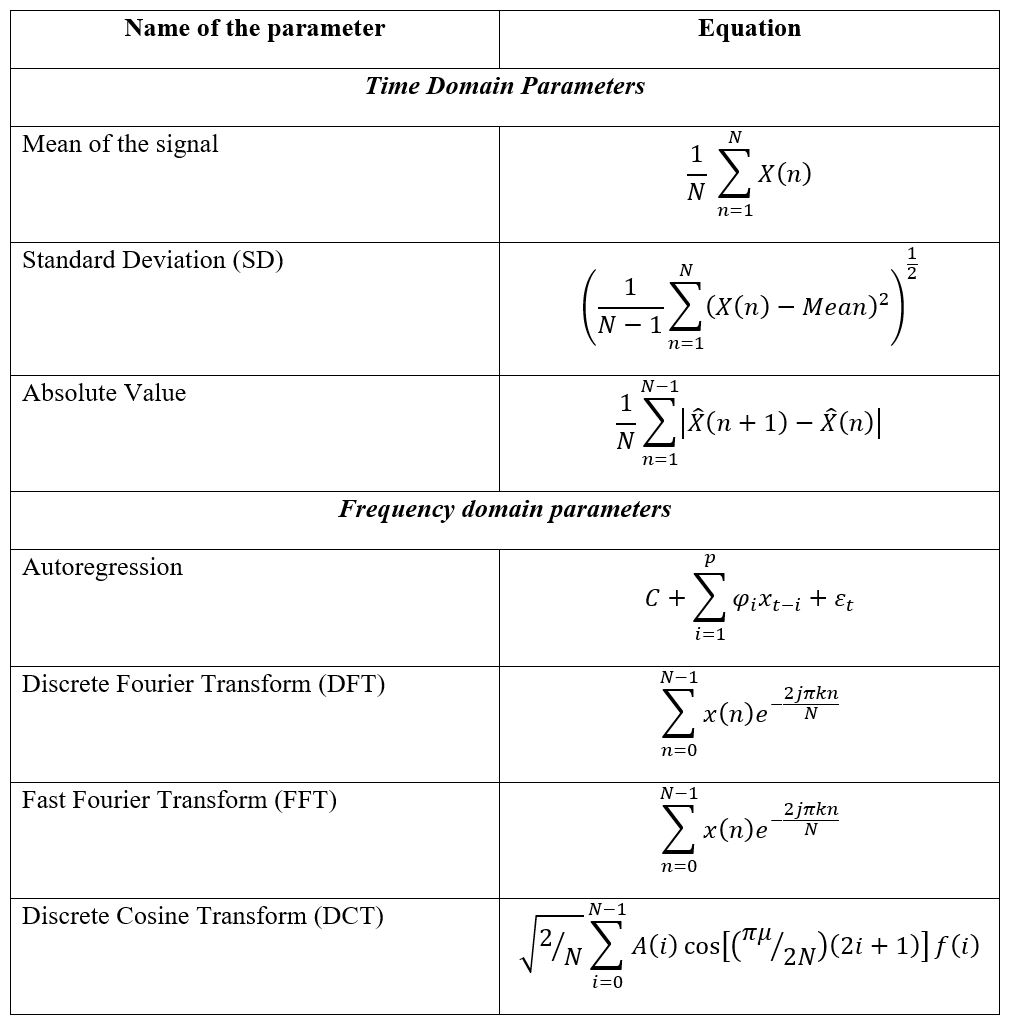}
\end{table}
\noindent
where $X(n)$ is the $n^{th}$ sample in the EEG signal, $\hat{X}(n) = \frac{X(n) - Mean}{SD}$, for $n=1, 2, ..., N$, $\varphi_i$ represents the auto-regression coefficients, x(t-i) is the previous value, and $\epsilon_t$ represents the error.      

\subsection{Classification}
Once the features are extracted, several classification methods and algorithms could be used in the Brain-computer Interface (BCI) systems. These include linear classifiers, non-linear classifiers, Naïve Bayesian classifiers, Neural Networks \cite{lotte2007review}, Support vector Machine (SVM), k-Nearest Neighbour (kNN) and others.

\subsubsection{Support Vector Machine (SVM)}
Support Vector Machine (SVM) is an example of supervised learning algorithm that is used for classifying the data. SVM is the oldest and traditional type of classification method that is used for classifying the Non-Invasive Brain signals. SVM is a preferred classification method because of its accuracy and low computation power. This classification method is used in identifying the Hyperplane in a space with several features. Mainly, SVM constructs a hyperplane as the decision surface in such a way that the margin of separation between positive and negative examples is maximized. The Hyperplane in a dimensional space is a plane which differentiates the different data points that are present in the space and these data points are referred to as the ‘Support Vectors’, who are responsible for the change in the position and orientation of the Hyperplane.
SVM is a linear classifier whose decision boundaries can be changed into non-linear by a small increase in the complexity using the kernel trick \cite{kolodziej2012linear}. SVMs are used mostly in biomedical applications because of their accuracy and easily can deal with the predictors \cite{essa23, libsvm, essa24, essa25}.

\subsubsection{Linear Discriminate Analysis (LDA)}
Linear Discrimination Analysis (LDA) is a popular classification and feature reduction technique that is mostly used while dealing with linear features/data. In the LDA, the linear data is separated into different classes with the help of hyperplanes as a separating plate. The data is separated equally with a co-variance matrix. LDA is a simple classifier that is easy to use \cite{kolodziej2012linear}, and it provides low computation \cite{abe2005support}. LDA provides better results on the linear data than compared to the non-linear EEG data.

\subsubsection{Bayesian Classifiers}
Bayesian classifiers are an example for conventional classifiers and work based on the probability theory \cite{hope2011workload}. Bayesian classifiers are popular in the pattern classification. Bayesian classifiers probability can be calculated with the help of a Bayesian rule, which can be given by the equation:

\begin{equation}
P(\omega|X) = \frac{p(x|\omega_i)P(\omega_i)}{p(x)}
\end{equation}
\noindent	
where 

\begin{equation}
p(x) = \sum_{i=1}^{n} p(x|\omega_i) P(\omega_i)                    
\end{equation}

\noindent
and $\omega$ is the random variable, i indicated the class and x is the random vector in an m-dimensional space. $ p(x|\omega_i)$ represents the conditional density \cite{lecun2015deep}.

\begin{figure}[h!]
	\centering
	\includegraphics[width=.85\textwidth]{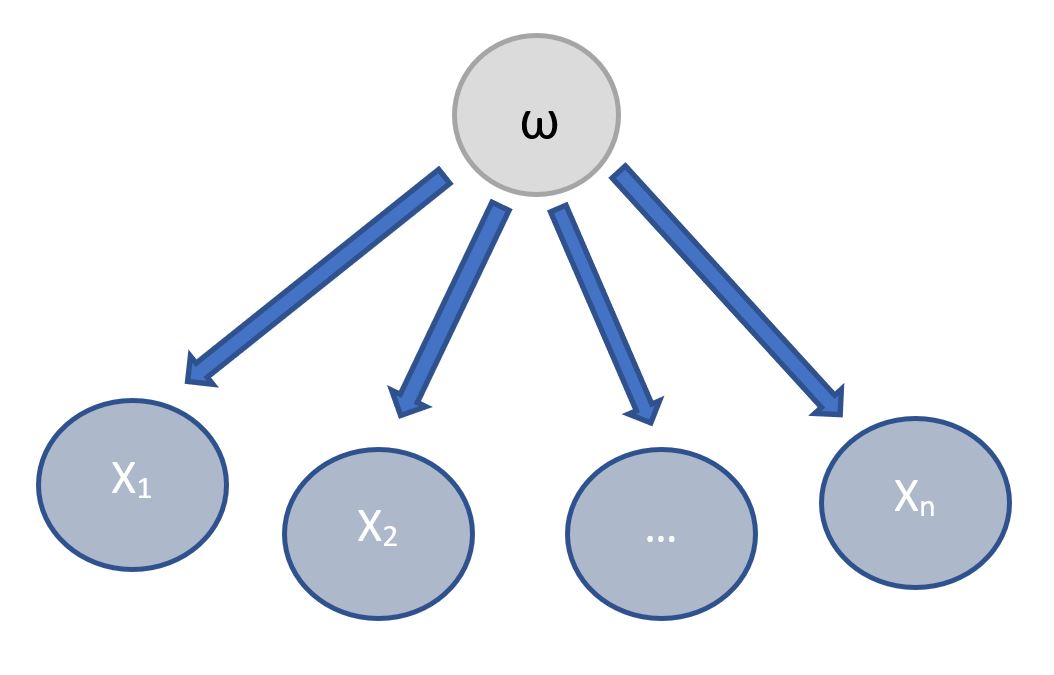}
	\caption{Bayesian Classifier.}
	\label{fig7}
\end{figure}

When Bayesian rule is used for the diagnostic problems, it includes classes that are both normal and abnormal. Therefore, a Bayesian classifier has a structure that has both classes and the feature nodes. In case of the EEG classification, the class nodes represent the workload conditions whereas the feature nodes represent the EEG frequency features. The Bayesian classifier is represented as shown in Fig. \ref{fig7}, where $\omega$ represents the class node (the classification) and x1, x2, …., xn represents the EEG frequency feature nodes.

\subsubsection{Machine Learning}
Machine learning (ML) is one of the subsets of Artificial Intelligence. ML is the study of different tasks by developing the mathematical models and algorithms, which are used for constructing computational models of artificial neurons. \cite{jordan2015machine}. ML techniques help in extracting the features from the trained data sets and develop machine learning models that try to mimic the behavior of the human brain to the improve the performance of a particular task without involving any human work \cite{mahmud2018applications}.

\subsubsection*{Artificial Neural Network (ANN)} 
\begin{figure}[h!]
	\centering
	\includegraphics[width=.9\textwidth]{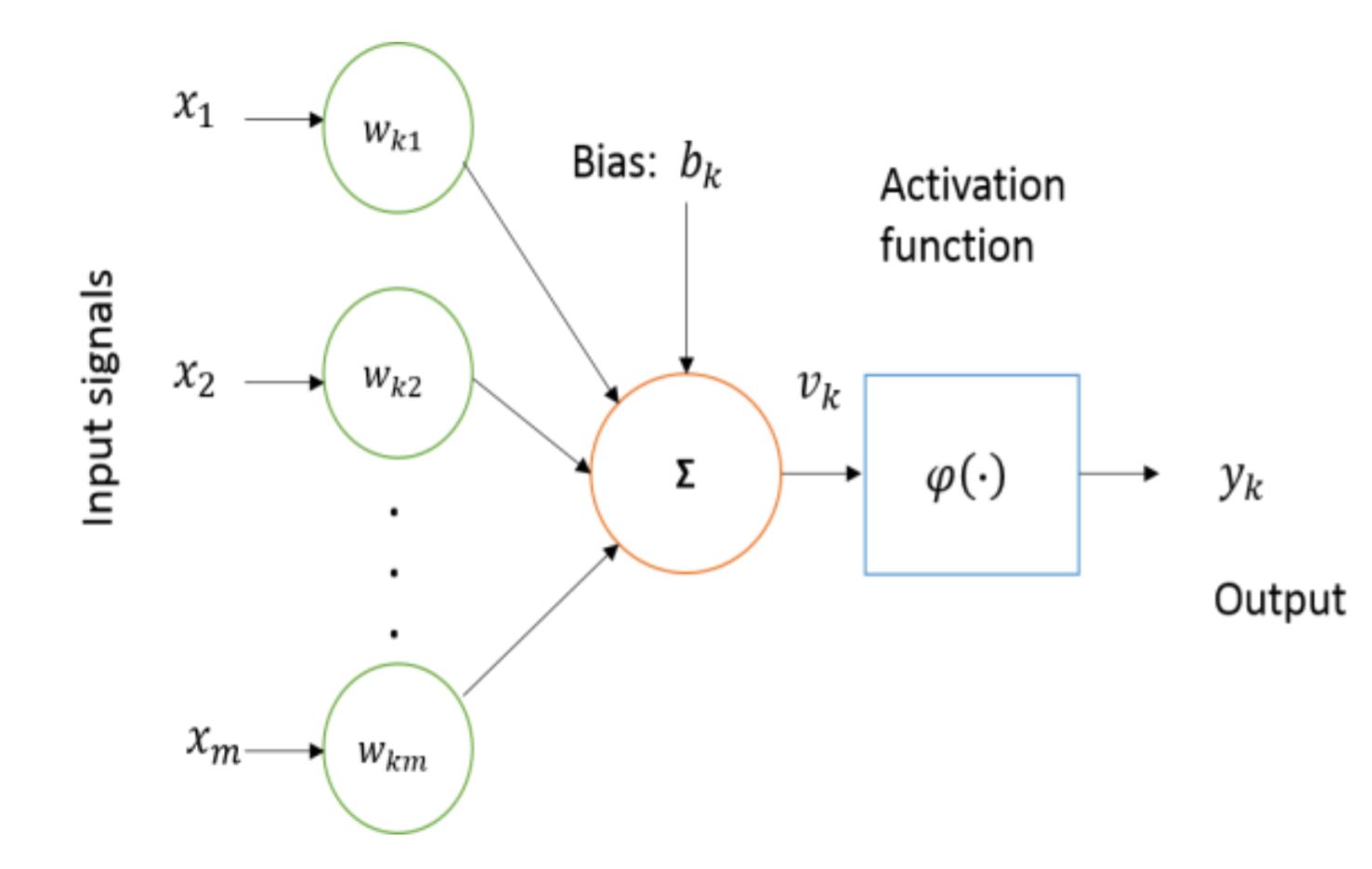}
	\caption{Single Layer Perceptron.}
	\label{fig8}
\end{figure}

Computer scientists have long been inspired by the human brain. The human brain can be described as a biological neural network which is an interconnected web of neurons transmitting elaborate patterns of electrical signals. The artificial neurons are the fundamental component for building ANNs. The inputs are received by neuron (node), which is the basic computational element, with the help of some internal parameters (weights and biases), which are learned during training and responsible for producing the outputs. The simplest possible example of one neural network that consists of a single neuron is called a perceptron. Fig \ref{fig8} shows the basic block diagram of a perceptron.  

\subsubsection*{Multilayer Perceptron (MLP)}
A multilayer perceptron also known as ‘Classic Neural Networks’ is a neural network of multiple layers that contains one input layer, one output layer, and one or more hidden layers. This network is connected to each other via one output layer to the other neurons input layer. In the MLP, the hidden layers play an important role of processing the input information and transmitting it to the output layer and keep the track of the accuracy and the performance of the system. The output from the multiple hidden layers is the result of the product of the weights of the hidden neurons and the input signals. The neural network includes the parameters like the training cycles with a learning rate of several nodes in each hidden layer that has an activation function associated with it \cite{larsen2011classification}.

When it comes to the EEG signal data, a number of frames of the EEG signal is obtained by averaging the spectrum using a window function which will be given as the input to the Multilayer perceptron (MLP). A study showed that the features that are extracted from the EEG signals are more accurate and clearer when the time-domain parameters are used comparing to the frequency-domain parameters. For example, using autoregression (AR) model as the feature extraction technique on the EEG signals of short-term intervals and giving its output as input to the MLP resulted in less accurate output when compared to the theoretically presented results. By processing the EEG data frame by frame, the MLP classification technique can be the neural network with generalization attributes. Table \ref{table3} describes the EEG data files that are classified with the original classes and the predicted classes that have been conducted experimentally and recorded by \cite{tsoi1994classification}. These experimental results present the classification technique used on the EEG data of the Schizophrenic and Obsessive-Compulsive Disorder (OCD) patients. 

\begin{table}[ht]
	\centering
	\caption{Classification performed on EEG data.} 
	\label{table3}
	\includegraphics[width=13cm,height=14.5cm]{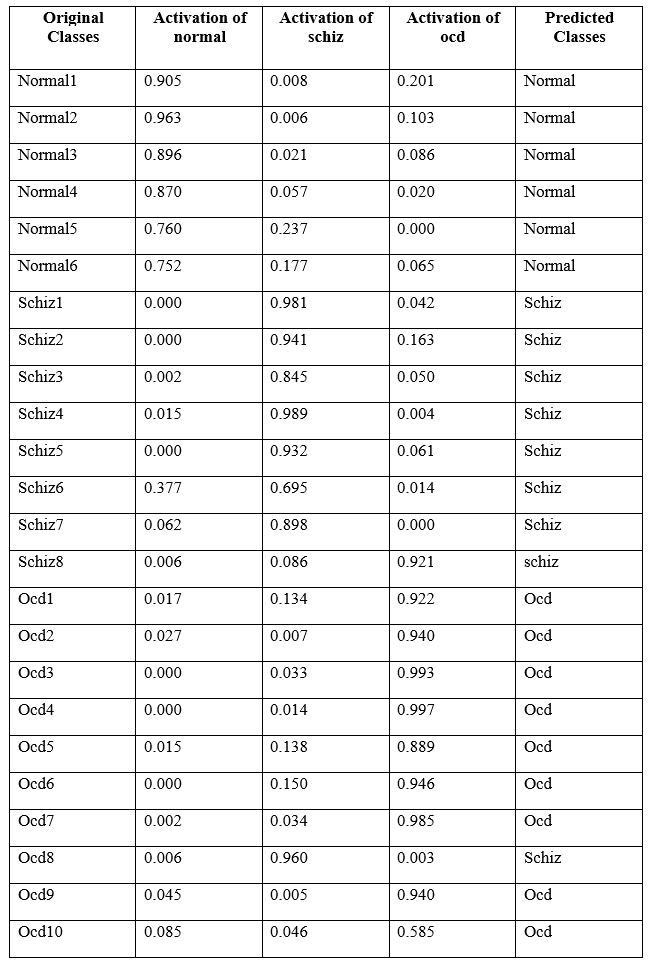}
\end{table}

\section{Deep Learning}
\begin{figure}[h!]
	\centering
	\includegraphics[width=.95\textwidth]{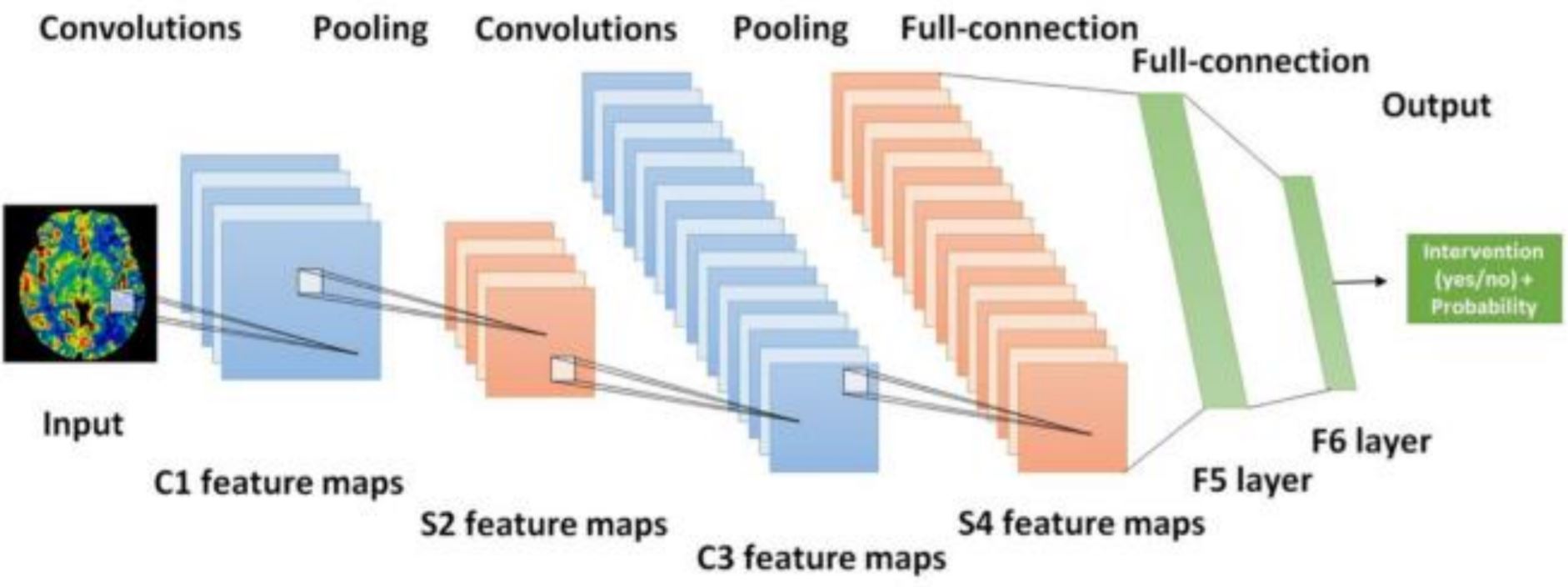}
	\caption{The overall architecture of the CNN includes an input layer, multiple alternating convolution and max-pooling layers, two fully-connected
		layers and one classification layer \cite{feng2018deep}.}
	\label{fig9}
\end{figure}

Deep learning (DL) is one of the subsets of Artificial Intelligence (AI) and class of Machine Learning (ML), which works similar to the human brain in terms of processing the data and creating the respective patterns that performing different activities, and it is this subfield that provides systems with the ability to learn and improve automatically from experience without being explicitly programmed. DL is used for making decisions based on the data that is being processed using neural networks \cite{lecun2015deep}, which have neural codes that let the data to be processed across a number of layers of the neural networks. Fig. \ref{fig9} shows the typical model of a Convolutional Neural Networks (CNNs) architecture \cite{feng2018deep}, which uses the temporal and spatial filters for classifying the EEG data that is acquired from the cortex region of the brain using electrodes. The samples that are collected are pre-processed using the artificial feature extraction techniques and then classification is done using a multi-layer CNN on the EEG features. The CNN model uses the training set, which has been trained initially and got used to the EEG features and there has been a reduction in the dimension. 

\begin{figure}[h!]
	\centering
	\includegraphics[width=.9\textwidth]{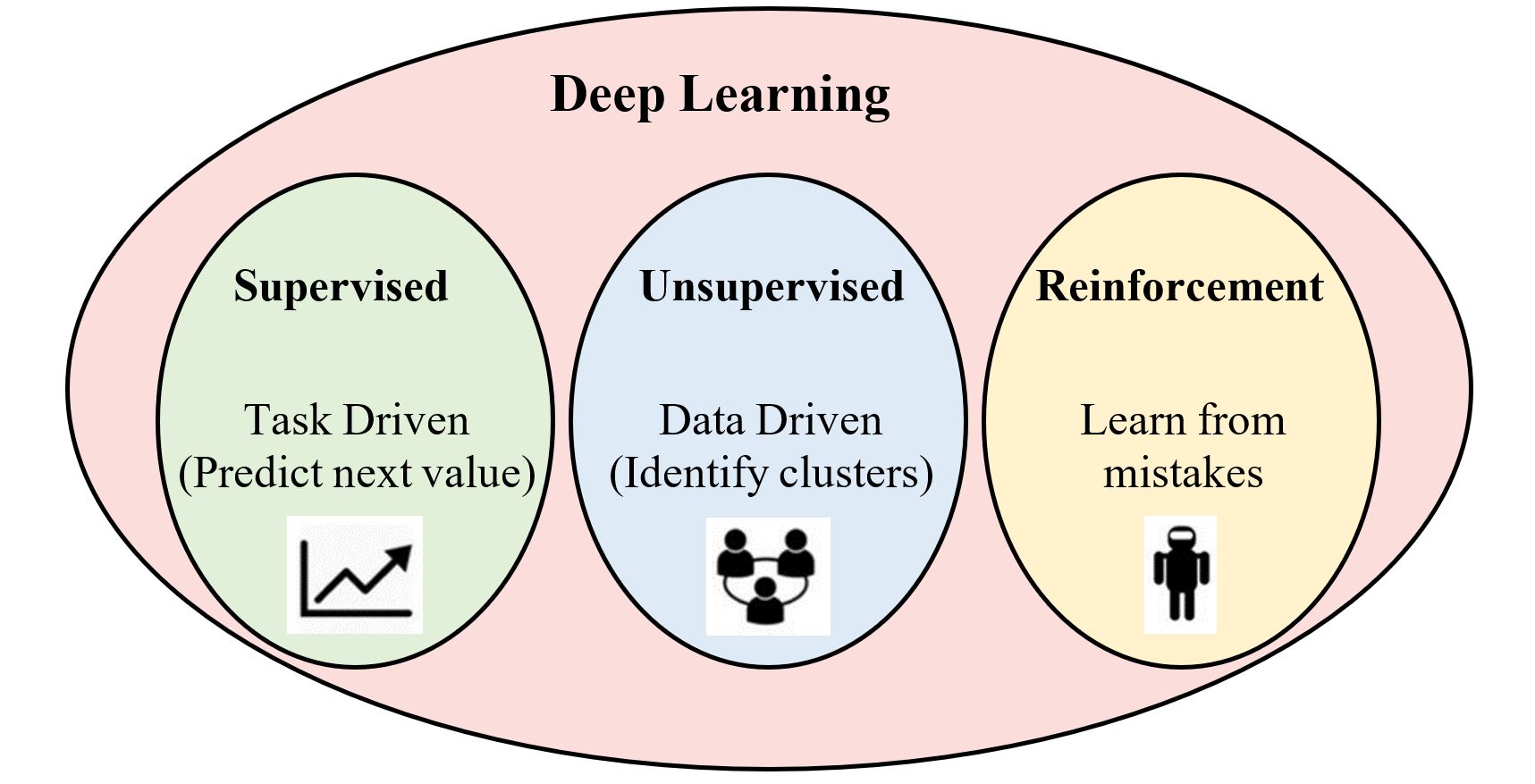}
	\caption{Category of deep leaning approaches.}
	\label{fig10}
\end{figure}

Deep learning approaches can be categorized to three main categories based on the aim of the techniques such as supervised learning, unsupervised learning, and reinforcement learning, which can be explained in Fig. \ref{fig10}.  

\subsection*{Supervised Learning Models}
Supervised learning is a learning technique that uses a specific training data to train the neural network (network circuit with a function). The function that is used in a supervised neural network model can be modified based on the output that is interpreted initially. The neural network’s function will be modified until a desired output or until it meets the targeted output. These supervised models are trained using the back-propagation algorithm. Some examples of several supervised deep learning methods are Convolutional Neural Networks (CNN), Deep Neural Networks (DNN), Recurrent Neural Networks (RNN), etc.  

\subsection*{Unsupervised Learning Models}
Unsupervised learning models are used in understanding and developing the predictive models for the unstructured or unlabeled data. Unsupervised learning models are not provided with an interpreted outputs unlike the supervised learning models. As there is no predicted or interpreted output, there is no training data set present in order to train the algorithm \cite{celebi2016unsupervised}. The algorithm is processed with the provided inputs resulting in an output. The conclusions are drawn from the resultant output of the model. The Auto Encoders (AE), generative adversarial network (GAN), and Restricted Boltzmann Machines (RBM) are examples of the unsupervised deep learning models.

\subsection*{Reinforcement Learning}
Reinforcement learning is the training of machine learning models to make a sequence of decisions for use in unknown environments. In this area of machine learning, the system interacts with a dynamic environment in which it must perform a certain goal using a straightforward loss function, which controls the convergence to the optimal action value function. Therefore, the system provides feedback in terms of rewards and punishments as it navigates its problem space. In this case, the human involvement is limited to changing the environment and adjusting the system of rewards and penalties.

\section{Recent Advances and New Frontiers of DL Techniques for Brain Signals Analysis}
Deep learning (DL) models have attracted attention in many areas for its superior performance and have contributed to various of brain signal applications. In this section, we will summarize the advanced studies and new frontiers on deep learning-based brain signals. 

\subsection{Motor Imagery Electroencephalography (MI-EEG)} 
The MI-EEG is a self-regulated EEG that can be detected by electrodes without an external stimulus. The authors in \cite{lun2020simplified} have used deep convolutional neural network (DCNN) for MI-EEG raw signals of nine pairs of symmetrical electrodes over the motor cortex region. In this research work, there were four different types of dataset used for the EEG data features extraction and model training, and four classification models were obtained. The CNN model used in \cite{lun2020simplified} has algorithms that improve the classification accuracy and probability of the trained deep CNN models based on their provided results as can be seen in Table \ref{table4} (global average accuracy of CNN models in different datasets) that is obtained from the cited paper. For more details, see the original paper \cite{lun2020simplified}.  


\begin{table}[ht]
	\centering
	\caption{CNN model global average accuracy in different dataset.} 
	\label{table4}
	\includegraphics[width=11cm,height=2.7cm]{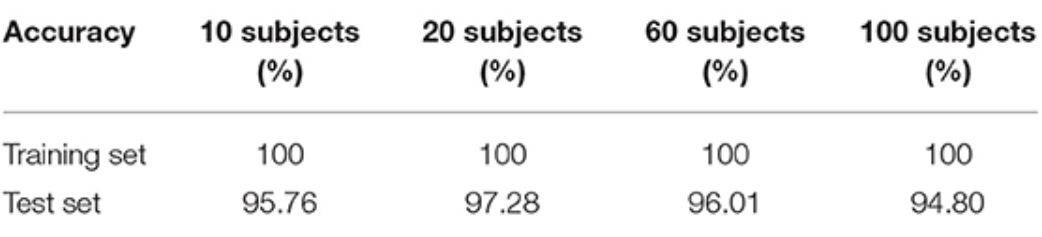}
\end{table}

There is another CNN method for EEG classification proposed by \cite{thomas2018eeg} which is based on the waveform-level to detect the abnormal EEG waveform that does not show a spike. In the CNN system proposed by \cite{thomas2018eeg} uses TensorFlow with a K40 Tesla GPU. The implemented CNN network uses different parameters such as the convolutional layers, filters, pooling layers, Rectified Linear Unit (ReLU) as the activation function, etc. The classifier uses 80\% of the data for training and 20\% of the data for validating the model. The EEG data that is used is folded into different datasets under different combinations. This folded data is used for training the model and getting predicted outputs. The output of the CNN system is mapped in the form of a [0,1] with an activation function. The higher value of [0,1], is the higher spikiness of the EEG waveform. The performance results of the EEG-waveform-level-CNN system are given by the area under curve (AUC) score which can be seen in Table \ref{table5} that is obtained from the cited paper. For more details, see the original paper \cite{thomas2018eeg}. 


\begin{table}[ht]
	\centering
	\caption{CNN model global average accuracy in different dataset.} 
	\label{table5}
	\includegraphics[width=12cm,height=6.5cm]{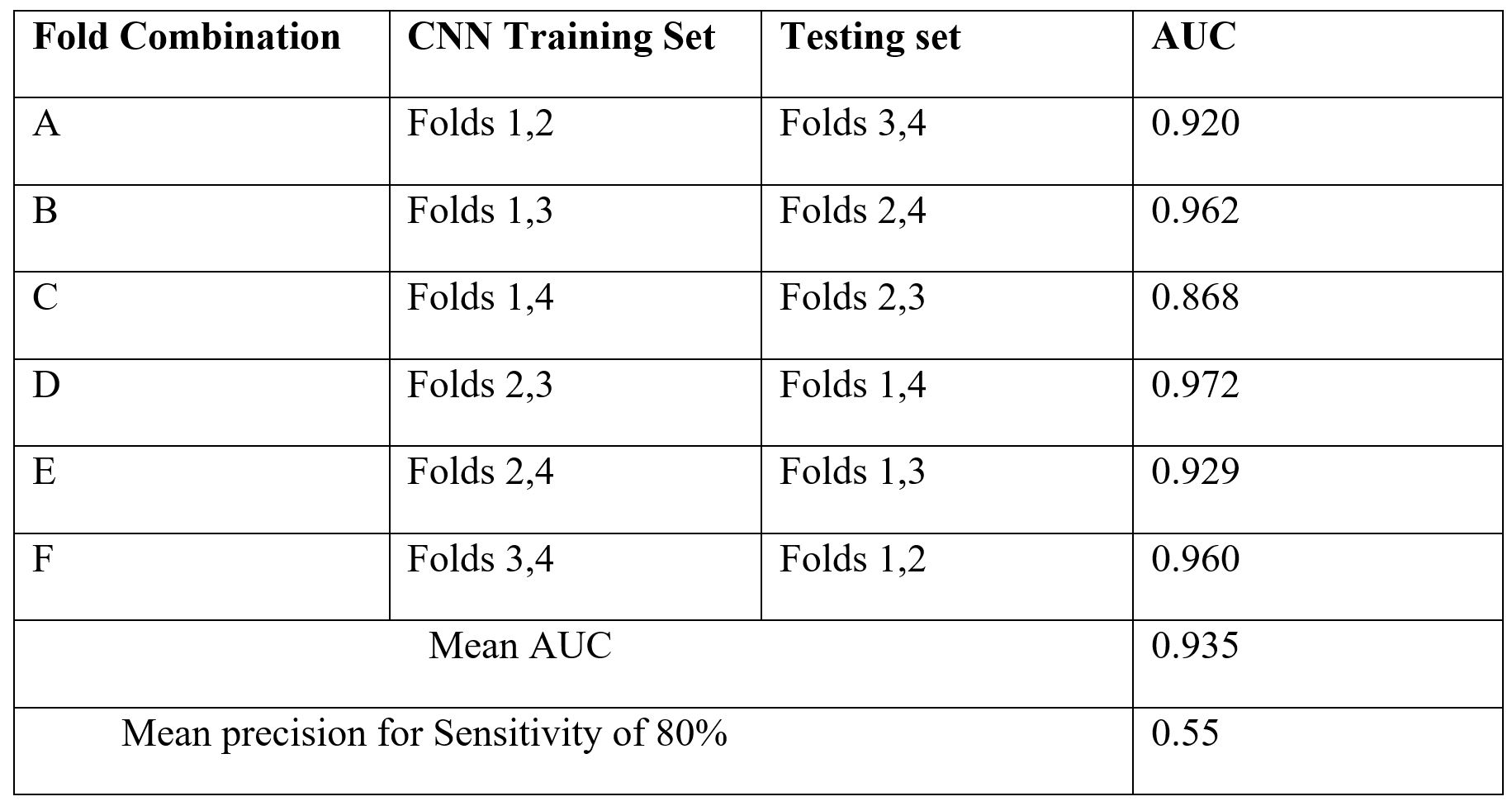}
\end{table}

\subsection{Interictal Epileptiform Discharges (IEDs)}
\begin{figure}[h!]
	\centering
	\includegraphics[width=.75\textwidth]{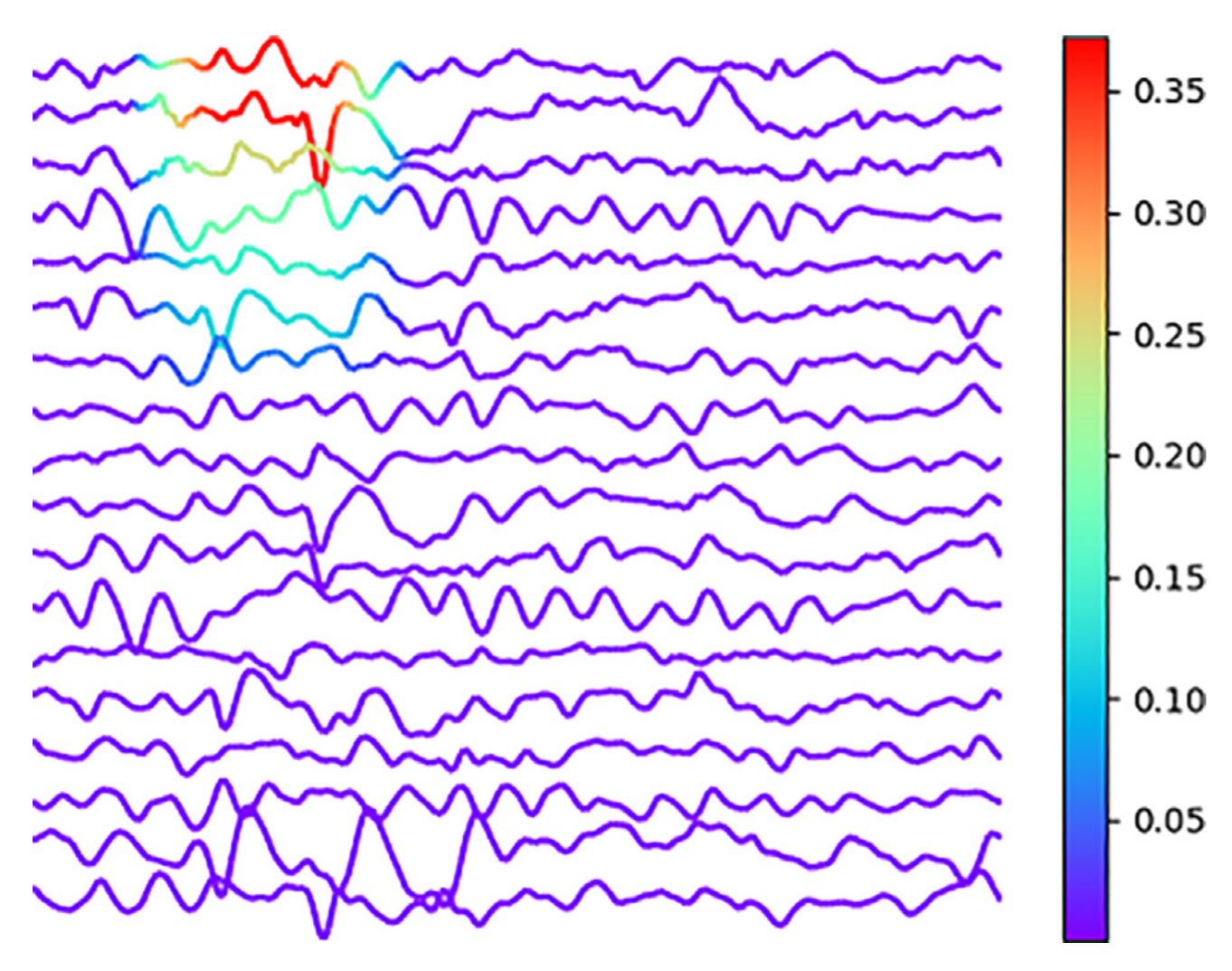}
	\caption{Likelihood heat-map obtained with the occlusion visualization technique on an EEG sample with a focal IED \cite{lourencco2019deep, da2021machine}. Warmer colors indicate higher importance for classification.}
	\label{fig11}
\end{figure}

The automated IEDs reflects an increased the probability of seizures and are routinely assessed by visual analysis of the EEG. There are several deep learning methods for IEDs detection that have been developed and proven their contribute to increase the interest in this field of research. While applying the deep convolutional neural network (DCNN) to raw EEGs was possible to achieve areas under the receiver operating characteristic curve (AUCs) higher than 0.9, indicative of algorithms with very good performance, the authors in \cite{jing2020development} have been developed and validated a DL method (SpikeNet) that is able to automatically detected IEDs and classified whole EEGs as IED-positive or IED-negative as the medical professionals do.     
In terms of the application of network visualization techniques, the authors in  \cite{lourencco2019deep, da2021machine} have been shown that their proposed DCNN network is able to automatically detect the IEDs in the samples and classifying them based on the same visual way that the medical professionals do, which can be seen in Fig \ref{fig11}. For more details, see \cite{lourencco2019deep, da2021machine}. 

\subsection{Ischemic Stroke Identification Based on EEG and EOG}
EEG rhythms is commonly used to diagnose the ischemic stroke, which are associated with the function of blood circulation around the brain as a carrier of oxygen to the brain. The authors in \cite{giri2016ischemic} have applied 1D Convolutional Neural Network (1DCNN) model that has the ability to distinguish the EEG and EOG stroke data from EEG and EOG control data. They have utilized several feature extraction techniques (24 handcrafted features) and compare their proposed deep learning method (100 and 200 Epoch, 5 times measurements each) with the conventional neural networks and naive bayes as we be seen in Table \ref{table6} that is obtained from the cited paper. For more details, see the original paper \cite{giri2016ischemic}. 

\begin{table}[ht]
	\centering
	\caption{Naive Bayes (NB), Neural Network (NN), and 1DCNN } 
	\label{table6}
	\includegraphics[width=12cm,height=6.4cm]{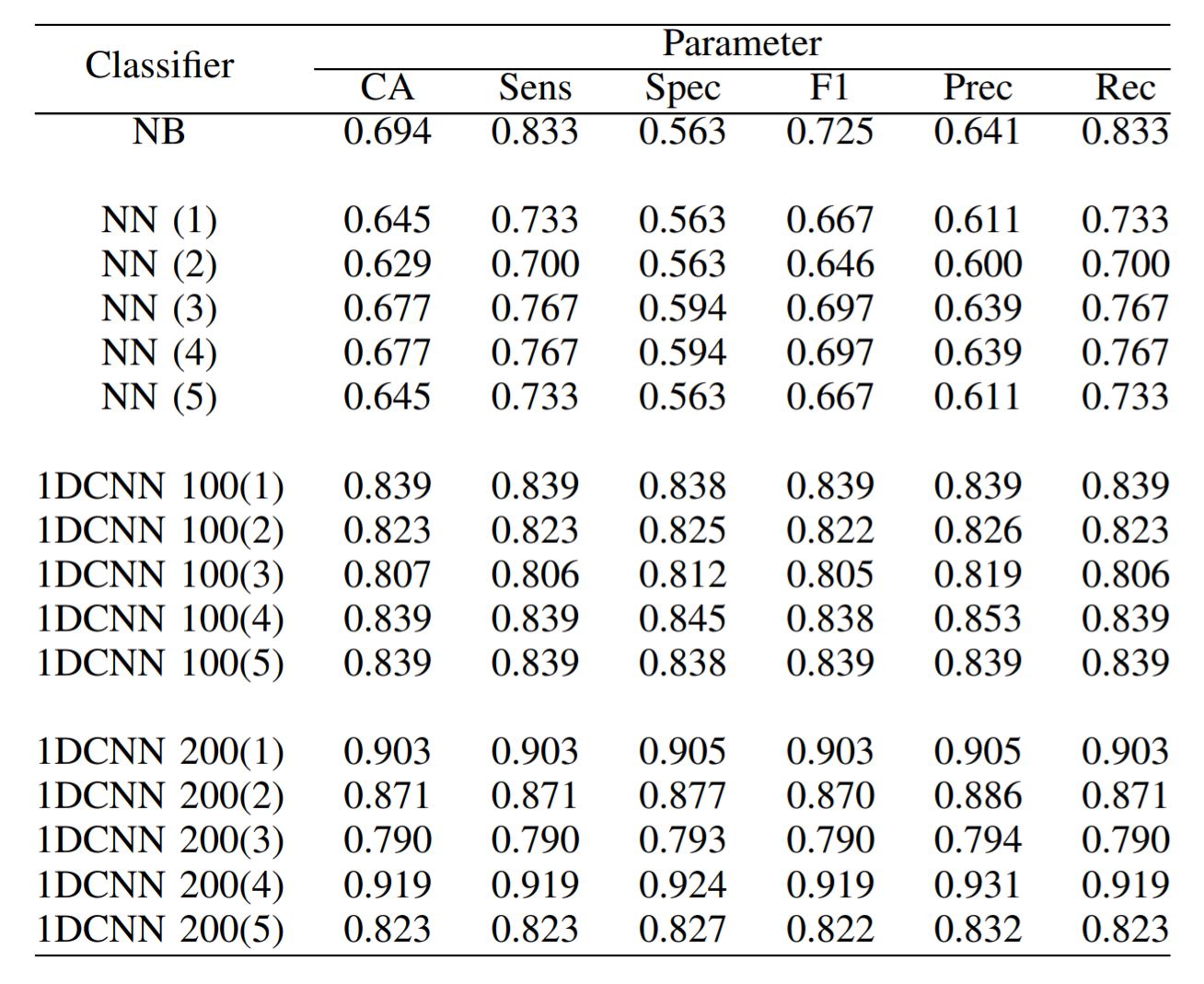}
\end{table} 	     
 
\section{Conclusion}
In this paper, the recent advances in deep learning models for non-invasive brain signal analysis have been summarized and presented. In order to provide simple guidelines and easy starting-up to help researchers to find the suitable deep learning algorithms for each category of the brain signals, we start with explaining the Brain-Computer Interface (BCI) system and dominant Deep Learning (DL) methods, followed by discussing the recent advances and new frontiers of DL techniques for brain signals analysis. 



\begin{backmatter}
%
%
%
%
%
Almabrok Essa is an Assistant College Lecturer at Cleveland State University. Cleveland, OH, USA. His research interests include artificial intelligence, computer vision, machine learning, pattern recognition, and remote sensing. He is a member of IEEE and SPIE.
%


\bibliographystyle{bmc-mathphys} 
\bibliography{bmc_article}      

\end{backmatter}
\end{document}